\documentclass[a4paper]{article}

\usepackage[squaren]{SIunits}
\usepackage{amssymb}
\usepackage[pdftex]{graphicx}
\usepackage{fancyhdr}
\usepackage{enumerate}
\usepackage{amsmath}
\usepackage{color}
\usepackage{setspace}
\usepackage[pdftex,bookmarks=true]{hyperref}
\usepackage{cite}
\usepackage{authblk}

\begin{document}

%\title{Measurement of angular resolution and polarisation asymmetry for $\gamma$ rays between 1.74 to 74\,MeV with the HARPO TPC}
\title{First measurement of polarisation asymmetry of a gamma-ray beam between 1.74 to 74\,MeV with the HARPO TPC}

\author[1]{P.~Gros\thanks{\noindent corresp. author: philippe.gros@llr.in2p3.fr}}
\author[2]{S.~Amano}
\author[3]{D.~Atti\'e}
\author[1]{D.~Bernard}
\author[1]{P.~Bruel}
\author[3]{D.~Calvet}
\author[3]{P.~Colas}
\author[4]{S.~Dat\'e}
\author[3]{A.~Delbart}
\author[1]{M.~Frotin}
\author[1]{Y.~Geerebaert}
\author[1]{B.~Giebels}
\author[5]{D.~G\"otz}
\author[2]{S.~Hashimoto}
\author[1]{D.~Horan}
\author[2]{T.~Kotaka}
\author[1]{M.~Louzir}
\author[2]{Y.~Minamiyama}
\author[2]{S.~Miyamoto}
\author[4]{H.~Ohkuma}
\author[1]{P.~Poilleux}
\author[1]{I.~Semeniouk}
\author[3]{P.~Sizun}
\author[2]{A.~Takemoto}
\author[2]{M.~Yamaguchi}
\author[1]{S.~Wang}

\affil[1]{LLR, \'Ecole Polytechnique, CNRS/IN2P3, Palaiseau, France}
\affil[2]{LASTI, University of Hy\^ogo, Japan}
\affil[3]{Irfu, CEA-Saclay, France}
\affil[4]{JASRI/SPring8, Japan}
\affil[5]{AIM, CEA/DSM-CNRS-Universit\'e Paris Diderot, France}
%\affil[6]{Irfu/Service d’Astrophysique, CEA-Saclay, France}

\pagestyle{plain}

\maketitle

\begin{abstract}
Current $\gamma$-ray telescopes suffer from a gap in sensitivity in the energy range between 100\,keV and 100\,MeV, and no polarisation measurement has ever been done on cosmic sources above 1~MeV.
Past and present e$^+$e$^-$ pair telescopes are limited at lower energies by the multiple scattering of electrons in passive tungsten converter plates. 
This results in low angular resolution, and, consequently, a drop in sensitivity to point sources below 1\,GeV.
The polarisation information, which is carried by the azimuthal angle of the conversion plane, is lost for the same reasons.

HARPO (Hermetic ARgon POlarimeter) is an R\&D program to characterise the operation of a gaseous detector (a Time Projection Chamber or TPC) as a high angular-resolution and sensitivity telescope and polarimeter for $\gamma$ rays from cosmic sources. 
It represents a first step towards a future space instrument in the \mega\electronvolt-\giga\electronvolt\ range.

We built and characterised a 30cm cubic demonstrator [SPIE 91441M], and put it in a polarised $\gamma$-ray beam at the NewSUBARU accelerator in Japan. 
Data were taken at photon energies from 1.74\,\mega\electronvolt\ to 74\,\mega\electronvolt\, and with different polarisation configurations. 

We describe the experimental setup in beam.
We then describe the software we developed to reconstruct the photon conversion events, with special focus on low energies.
We also describe the thorough simulation of the detector used to compare results.
Finally we will present the performance of the detector as extracted from this analysis and preliminary measurements of the polarisation asymmetry.% and of the angular resolution.

This beam-test qualification of a gas TPC prototype in a $\gamma$-ray beam could open the way to high-performance $\gamma$-ray astronomy and polarimetry in the MeV-GeV energy range in the near future. 
\end{abstract}

\section{Introduction}

%$\gamma$-ray astronomy provides insight into understanding the non-thermal emission of some of the most violent objects in the Universe, such as pulsars, active galactic nuclei (AGN) and $\gamma$-ray bursts (GRBs), and thereby into understanding the detailed nature of these objects.
$\gamma$ rays are produced in some of the most violent objects in the Universe, such as pulsars, active galactic nuclei (AGN) and $\gamma$-ray bursts (GRBs).
The field of $\gamma$-ray astronomy is concerned with the study of these non-thermal emissions, thus allowing us to gain an understanding of the detailed nature of these extreme objects.

Below $\approx$1\,\mega\electronvolt\, Compton telescopes ($\gamma e^- \rightarrow \gamma e^-$) are highly performant, but the Compton cross-section, and therefore the telescopes sensitivity, decreases with photon energy.
Above $\approx$1\,\giga\electronvolt\, pair telescopes ($\gamma Z \rightarrow Z e^+ e^-$) are highly performant, but they use high-Z converter plates (tungsten in Fermi's Large Area Telescope, {\it Fermi}-LAT) which severely degrade their resolution at low energy.
%The gap in resolution and sensitivity in the \mega\electronvolt-\giga\electronvolt\ range hinders the observation and the understanding of GRBs, whose spectra mostly peak in the MeV region.
%It could also bias the description of the Blazar sequence.
The gap in resolution and sensitivity in the \mega\electronvolt-\giga\electronvolt\ range hinders the study of many broad-band astrophysical emitters.
This includes for instance
% GRBs, whose spectra often peak in the MeV region, and 
blazars, which are postulated to comprise a sequence, the so-called blazar sequence~\cite{}, whose properties could be biased by the absence of sensitive observations in the \mega\electronvolt-\giga\electronvolt\ range.
It is also difficult to distinguish sources in crowded regions of the sky, such at the galactic plane.
To a large extent, the \mega\electronvolt-\giga\electronvolt\ sensitivity gap~\cite{Schoenfelder} is an angular resolution issue.
The angular resolution of pair telescopes can be improved, from the {\it Fermi}-LAT's $\approx 5\degree$ at $100\,\mega\electronvolt$~\cite{Ackermann:2012kna} to $1 - 2 \degree$, by the use of pure silicon trackers, i.e. without any tungsten converter plates~\cite{ASTROGAM,PANGU,Compair}. 

\begin{figure} [th]
 \begin{center} 
 \includegraphics[width=0.65\linewidth]{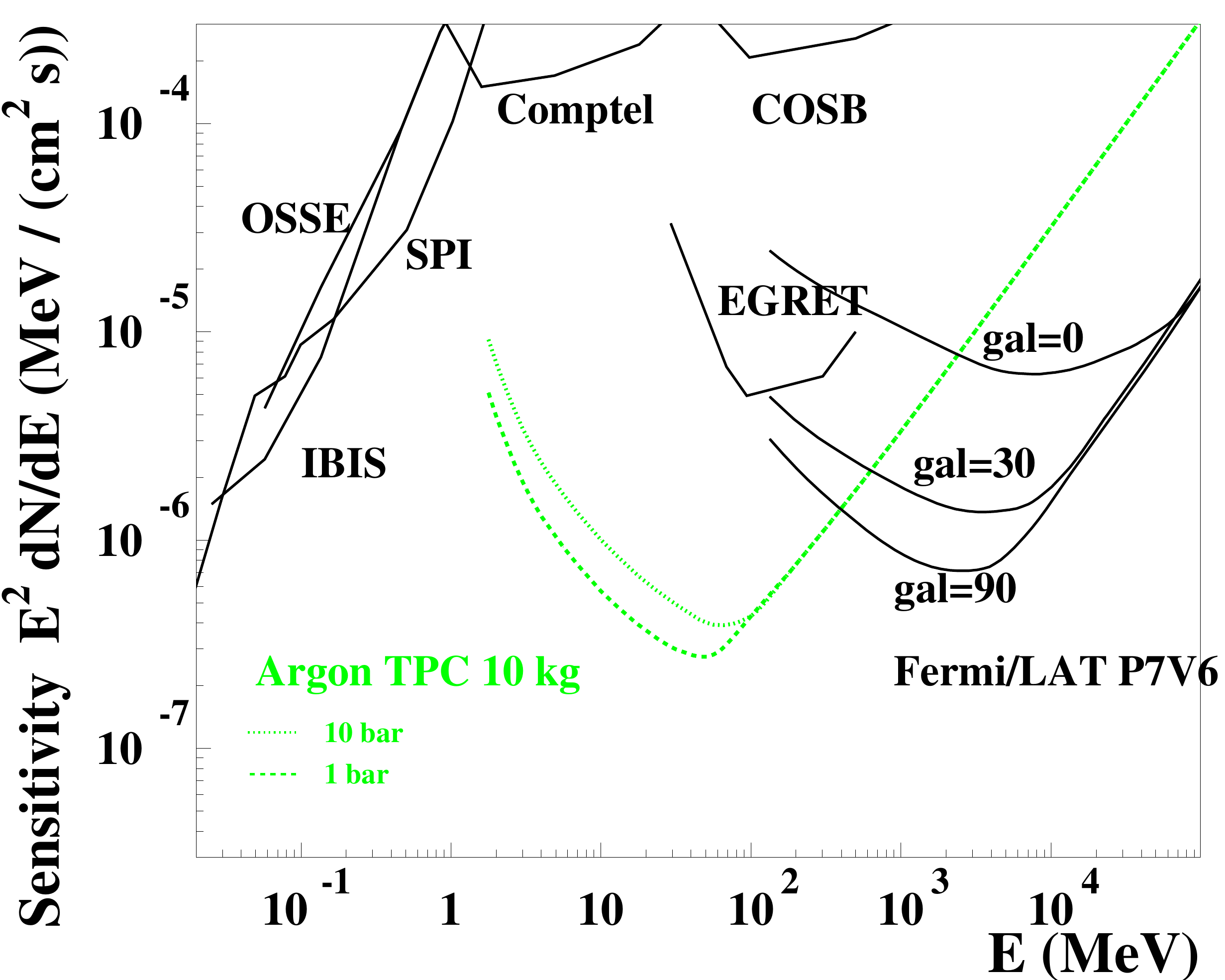}
\caption{
\label{fig:sensitivity}
 Differential sensitivity as a function of energy
 (argon-gas-based HARPO TPC, green) compared to the $90\degree$ galactic latitude
 performance of the {\it Fermi}-LAT \cite{Ackermann:2012kna} and of the
 Compton telescope COMPTEL \cite{Schoenfelder}.
 Adapted from \cite{Bernard:2012uf}.
}
\end{center}
\end{figure}

An even better resolution of $\approx 0.4\degree$ can, however, be obtained with a gaseous detector such as a time-projection chamber (TPC).
Together with the development of high-performance Compton telescope, filling the sensitivity gap for point-like sources at a level of $\approx 10^{-6}\,\mega\electronvolt\per(\centi\meter^2 \second)$ is within reach (see Ref.~\cite{Bernard:2012uf} and Fig.~\ref{fig:sensitivity}).

Furthermore, the use of a low density converter-tracker, such as a gaseous detector, enables the measurement of the polarisation fraction before multiple scattering ruins the azimuthal information carried by the pair \cite{Bernard:2013jea}.
Since the sensitivity of the Compton process to polarisation decreases as $1/E_{\gamma}$~\cite{Bernard:2013jea}, this is the only possibility for high-energy polarimetry.
The polarisation of sources was never available for $\gamma$-rays above $1\,\mega\electronvolt$ for reasons explained in Ref.~\cite{Mattox}.
At lower energies, in the radio to X-ray energy range, the measurement of the linear polarisation of the emission is a powerful tool for understanding the characteristics of cosmic sources.
It gives us access to information on the magnetic-field structure not otherwise readily available.
Extending the measurement to $\gamma$ rays will open new levels of understanding of these sources.

$\gamma$-ray polarimetry would, for example, provide insight into the understanding of the value and turbulence of magnetic fields in the $\gamma$-ray emitting jet structures of most $\gamma$-ray emitting sources. 
It could enable us to distinguish between the leptonic and hadronic nature of the emitting particles in blazars~\cite{Zhang:2013bna}, a long-standing open question in AGN research.

\section{The HARPO Experimental setup}

The HARPO (Hermetic ARgon POlarimeter) detector is a demonstrator of the performance of a TPC for measuring polarised $\gamma$ rays.
It was designed for a validation on the ground, in a photon beam.
The most critical constraints related to potential space operation were taken into account, such as the reduced number of electronic channels, and long-term gas quality preservation.

\subsection{The HARPO TPC}

\begin{figure} [ht]
  \begin{center}
    \includegraphics[width=0.5\textwidth]{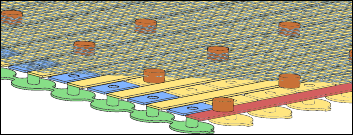}
  \end{center}
  \caption{
    \label{fig:exp:strips} 
    Detail of the micromegas and readout strips configuration.
  }
\end{figure}

The HARPO detector~\cite{Gros:TIPP:2014} is a 30\,\centi\metre\ cubic TPC, filled with Ar:isobutane (95:5) gas mixture up to 5\,\bbar.
It is equipped with a drift cage providing a 220\,\volt\per\centi\metre\ drift field.
The electrons produced by ionisation of the gas drift along the electric field toward the readout plane at a constant velocity $v_{drift} \approx 3.3\,\centi\metre\per\micro\second$.
The arrival time can therefore be converted to the $Z$-coordinate with the formula:
\begin{equation}
\label{eq:drift}
Z = v_{drift}(t_{signal}-t_{0})
\end{equation}
The readout plane is equipped with two Gas Electron Multipliers (GEMs) and one Micromesh Gas Structure (Micromegas) to amplify the electrons.
The amplified electrons' signal is collected by two sets of perpendicular strips (regular strips in the $X$-direction, and pads connected together by an underlying strip in the $Y$-direction, see Fig.~\ref{fig:exp:strips}).
The signals from the strips are read out and digitised with a set of AFTER chips~\cite{Calvet2014zva} and associate Front End Cards (FECs).

\begin{figure} [ht]
  \begin{center}
    \begin{tabular}{ccc}
      \includegraphics[width=0.3\textwidth]{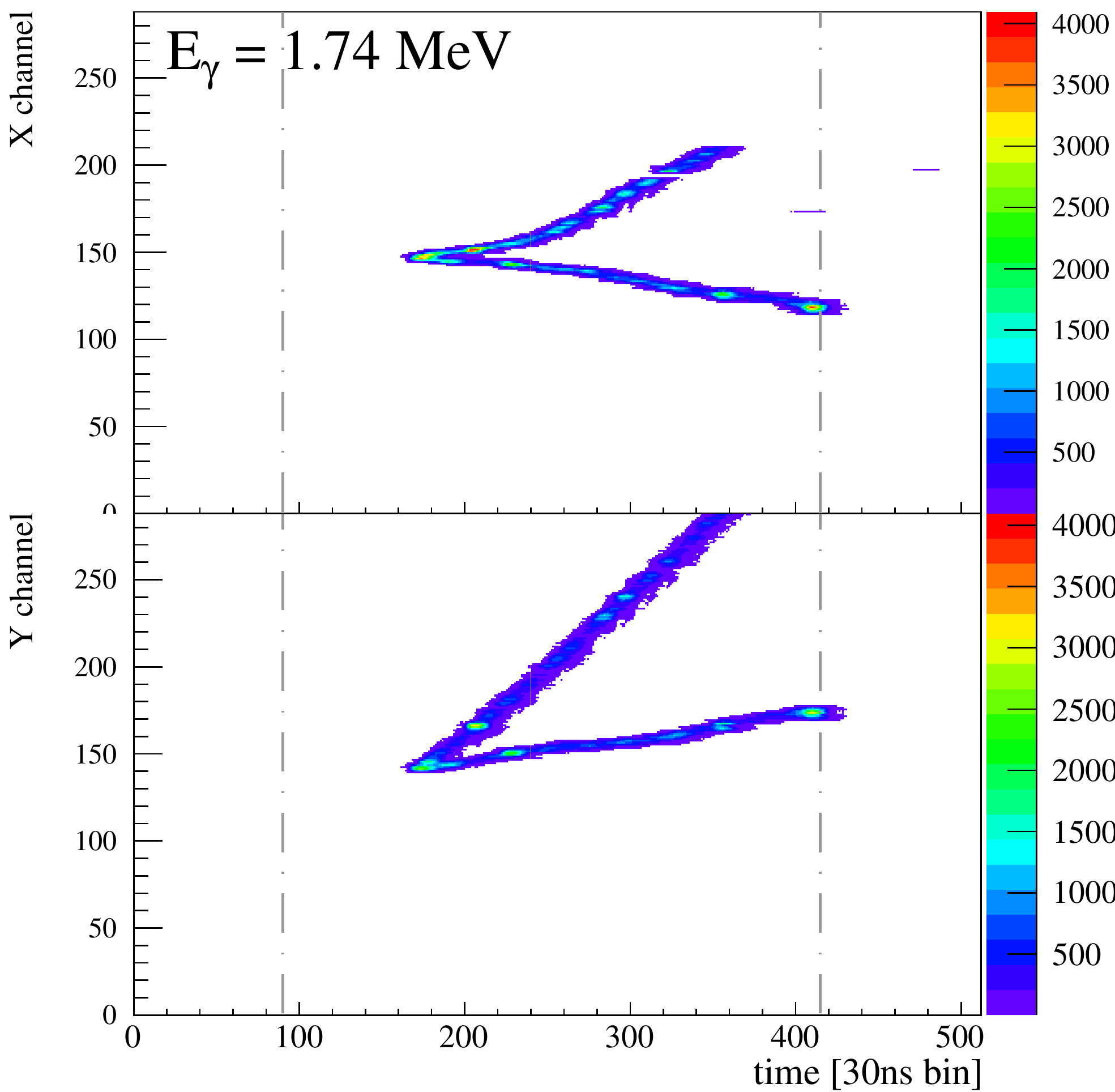} &
      \includegraphics[width=0.3\textwidth]{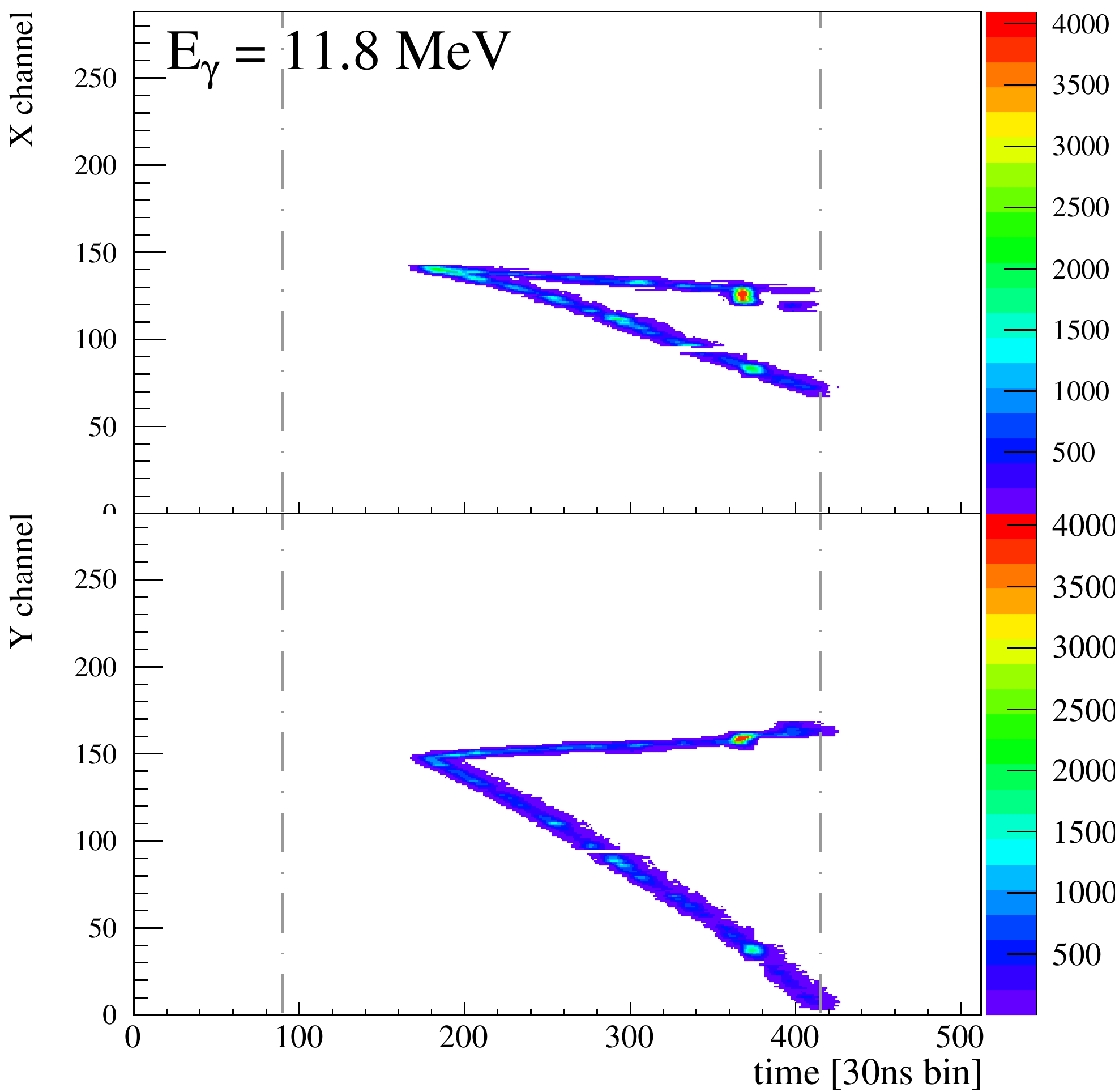} &
      \includegraphics[width=0.3\textwidth]{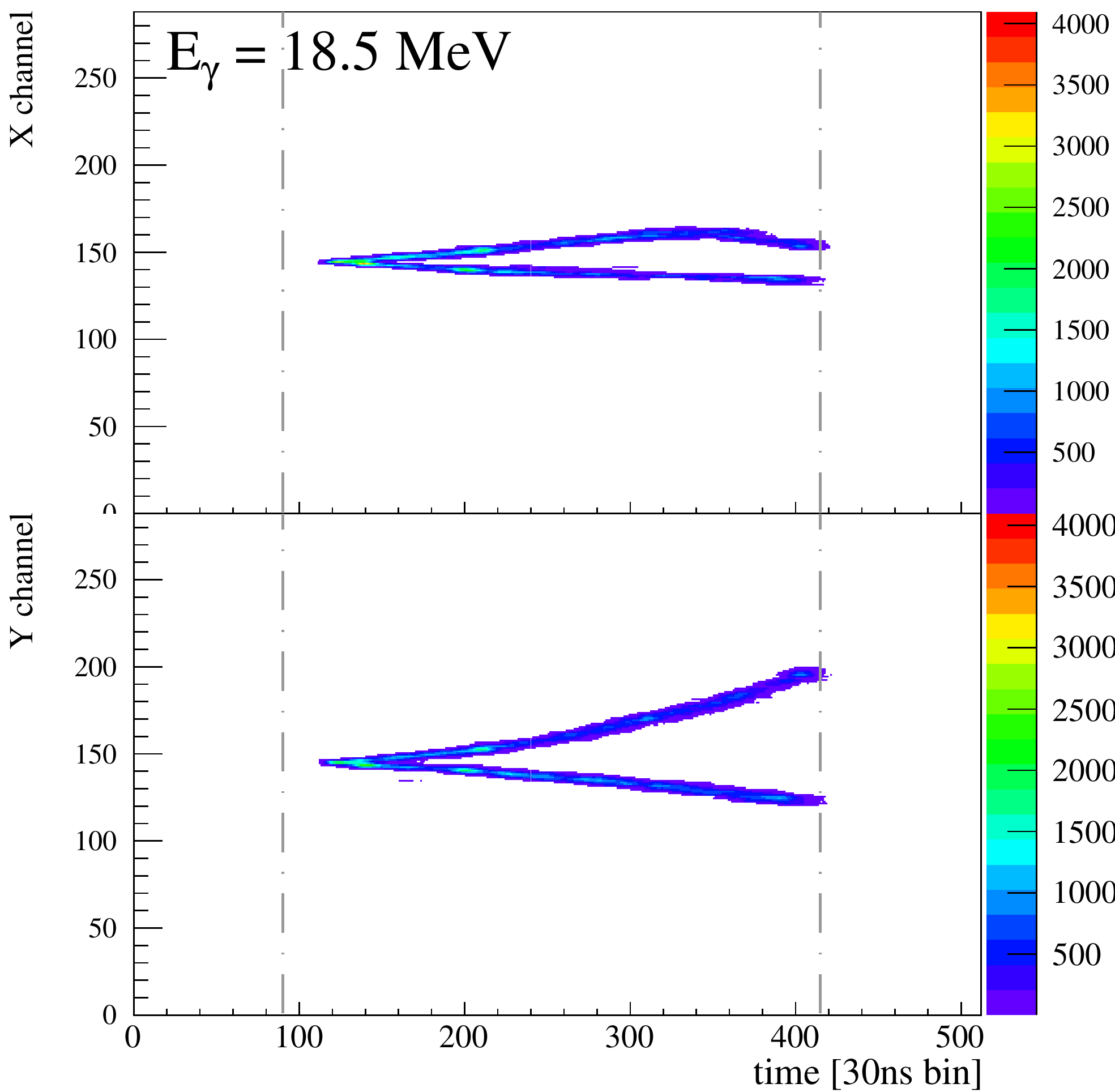} \\
      \includegraphics[width=0.3\textwidth]{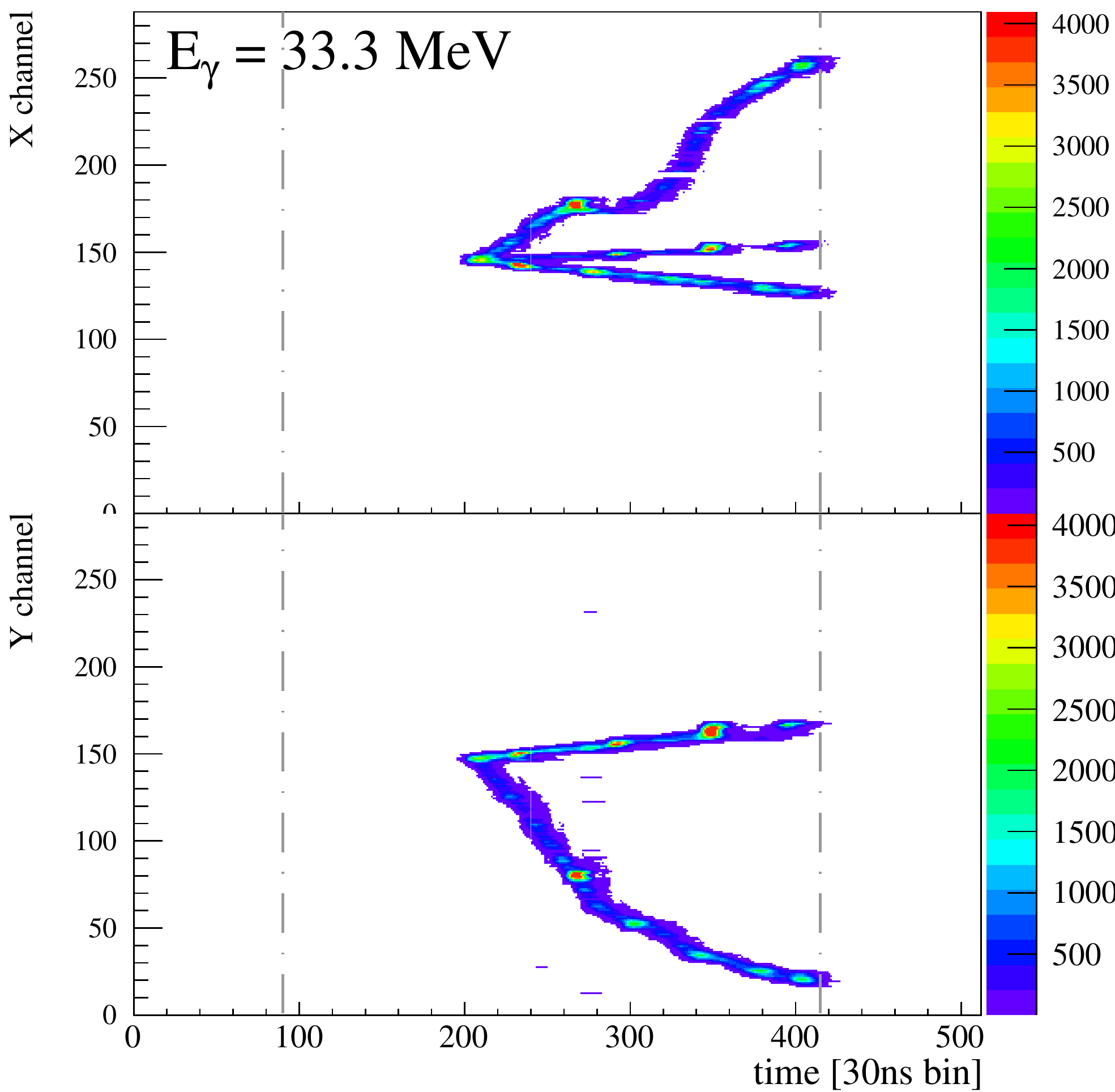} &
      \includegraphics[width=0.3\textwidth]{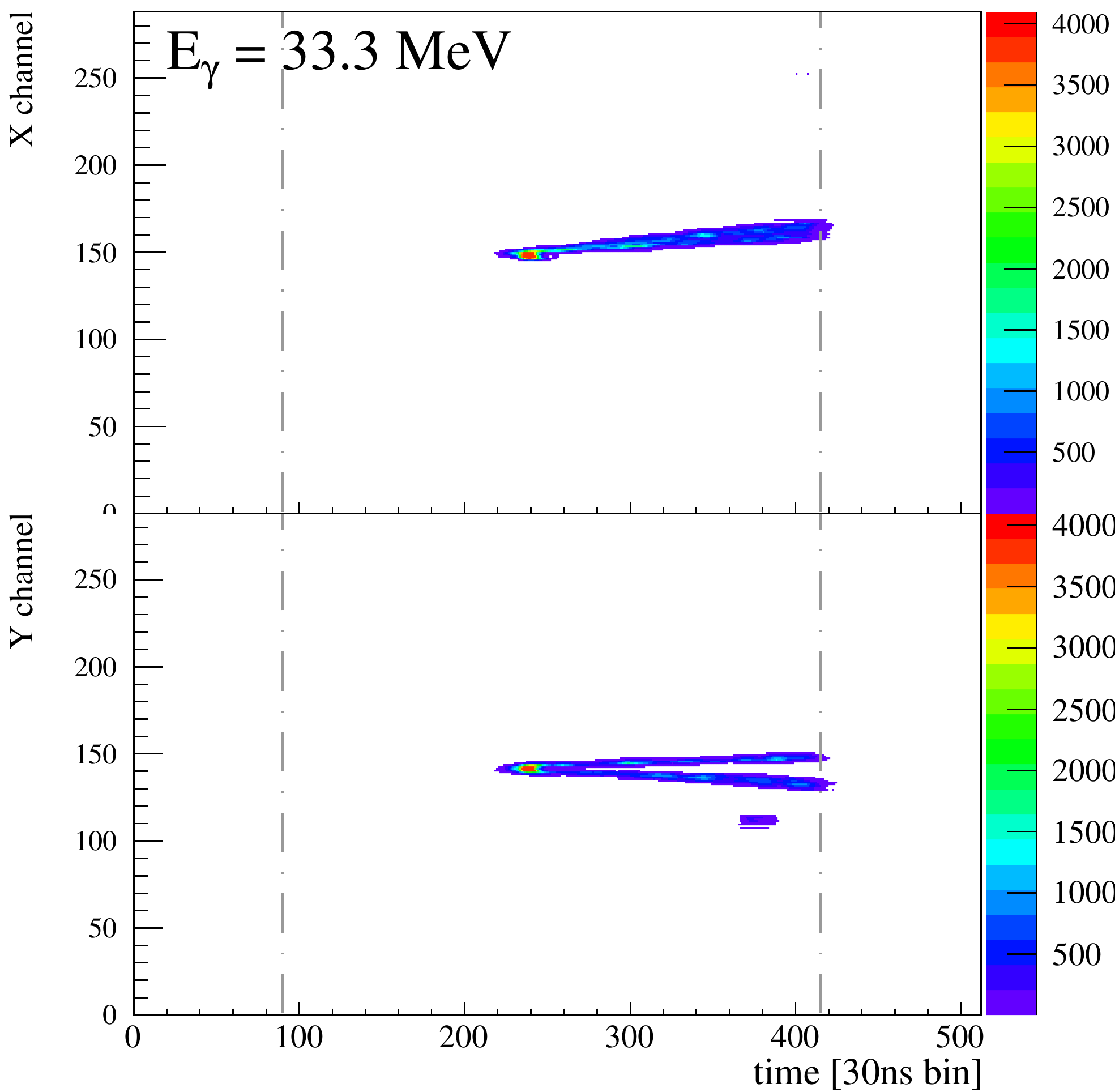} &
      \includegraphics[width=0.3\textwidth]{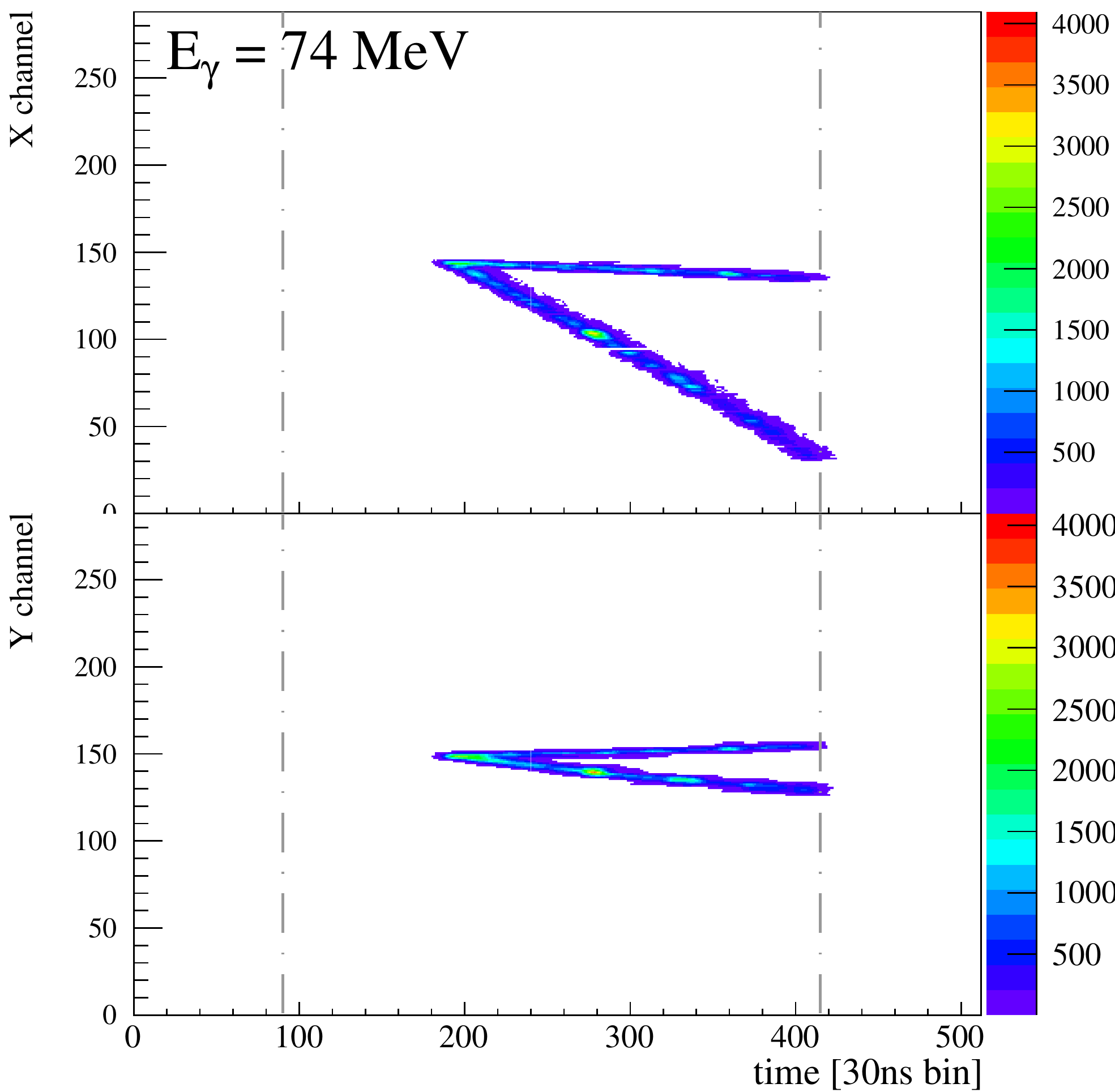}
    \end{tabular}
  \end{center}
  \caption{
    \label{fig:exp:event} 
    Typical conversion events in the HARPO detector, for different photon energies. 
    The top and bottom parts of the images show the two projections $XZ$ and $YZ$ (where $Z$ is proportional to the drift time of the electrons in the TPC).
    We can see that at higher energies, there is less scattering, and smaller opening angles.
%    A 16\,\mega\electronvolt\ photon converted into electron-positron pair.
    On the top left we see an asymmetry in the track directions, although the photon beam is in the $Z$ direction. 
    This is because part of the momentum is taken by the undetected recoiling nucleus.
    In the bottom left, we see a triplet conversion.
    In the bottom middle, we see that in the top projection ($XZ$) the two tracks are overlapping.
  }
\end{figure}

The signal recorded for an event in the TPC comprises the $XZ$ and $YZ$ projections of the deposited charge in the gas, where the $Z$ direction corresponds to the drift time of the electrons.
Several examples of pair-conversion events in the TPC are shown in Fig.~\ref{fig:exp:event}.

\subsection{Data taking at the NewSUBARU gamma-ray beam line}

The HARPO TPC was set up in the NewSUBARU polarised photon beam line~\cite{Horikawa2010209}.
The photon beam is produced by Inverse Compton Scattering (ICS) of an optical laser on a high energy (0.6-1.5\,\giga\electronvolt) electron beam.
Using lasers of various wavelengths and different beam energies (as described in more detail in~\cite{Wang:TPC:2015}), we scanned 13 photon energies from 1.74\,\mega\electronvolt\ to 74\,\mega\electronvolt.
%Due to complications with one of the lasers, the work presented here focuses on just 9 energy points, starting at 4.68\,\mega\electronvolt.

In order to minimise systematic effects due to the geometry of the detector, the detector itself was rotated around the beam axis to 4 different angular positions (-45, 0, 45 and 90\,\degree).
Finally, for some configurations (in particular at low energy), data were also taken with randomly polarised photons.

Most data were taken at a TPC pressure of 2.1\,\bbar.
A short scan of pressure up to 4.0\,\bbar\ was also done in the beam.

\subsection{Electronics saturation}

After analysis we found that the charge preamplifiers of the AFTER chips were often saturated during the data taking in beam.
If too much charge was accumulated on a single channel (which was the case in particular when a track was aligned with the drift direction), the preamplifier stopped responding linearly, and quickly stopped giving any signal at all.
The effects on our data were amplified by the beam configuration, which accumulated the signal on a few channels.
They were also made worse by the high luminosity.
This was reduced by changing the detector alignment, but it still affects a large fraction of the beam data.
Fortunately, most of the pair conversion information is contained in the early part of the event, before the saturation.
The reconstruction was therefore adapted to minimise the effect on the final results.

This saturation can be greatly and easily reduced by changing the dynamic range in the programmable electronics.
This should not affect the performance of the detector because the charge resolution is not critical for the reconstruction.
This was unfortunately not known at the time of the data taking in beam.
This effect will have very little impact on the performance for measuring cosmic photons that have random directions and a lower rate.

\section{Simulation of the HARPO detector}

We developed a complete simulation to describe the response of the HARPO detector.
It contains three main components: an event generator describing the conversion of photons in the gas, a Geant4 based simulation of the interaction of the electron-positron pair with the gas, and a custom description of the processes and geometry of the TPC.

\subsection{Photon conversion event generator}

There was no event generator available for photon conversion to $e^+e^-$ pair~\cite{Hirayama:2005zm,Salvat:2015lni,Geant4,Iparraguirre:2011zz} that fulfils the following requirements:
\begin{itemize}
\item exact down to threshold, that is, without any low-energy approximations;
\item complete, that is, that provides a sampling of the five-dimensional differential cross section;
\item polarised.
\end{itemize}
We therefore developed such a tool and characterised it with respect to expressions for 1D projection calculations published in the past (section 3 of Ref.~\cite{Bernard:2013jea}). 

\subsection{Electron propagation and ionisation with Geant4}

The position and momentum of the electron and positron are used as input for a simple simulation using Geant4~\cite{Geant4}.
Only the gas volume of the TPC is described in this simulation.
The electrons are propagated through the gas, taking into account, in particular, multiple scattering and gas ionisation.
This simulation provides the position of ionisation interactions in the gas volume, as well as the number of ionisation electrons.

\subsection{TPC response}

Since Geant4 cannot describe the electron drift and electronics response, we developed a standalone simulation of the HARPO TPC based on work done for the ALICE TPC~\cite{Christiansen2009149} and LCTPC~\cite{eudetmemo201036}.
This simulation takes as input a list of the ionisation electrons with their position (provided by Geant4).
For each individual electron, it applies a drift through the gas, taking into account drift velocity, Gaussian diffusion, and potential capture.
All the parameters are calibrated from the data, and are found to be consistent with values obtained using simulations with Magboltz~\cite{Biagi1999234}.

The amplification and readout of the electrons is described with a simple exponential gain fluctuation and Gaussian spatial spread.
The actual detector geometry is used to segment the signal.
The gain and space response are also calibrated using experimental data.

Finally, the electronics time response including shaping, noise and digitisation is added.
The saturation effects mentioned above, which noticeably affected our data, are also implemented.

The simulation output is in the same format as the real data, and all the reconstruction and analysis are applied identically to both.

\section{Reconstruction of photon conversions into e$^+$e$^-$ pairs in HARPO}

The diffusion of the signal, together with the large scattering of the particles in the gas, make it very difficult to use conventional tracking methods.
The large multiple scattering prevents the use of straight line finding algorithms such as Hough transforms~\cite{Hough}.
%On the other hand, the spread of the signal with high correlation makes it very difficult to use a Kalman filter~\cite{Fruhwirth:1987fm} to take into account multiple scattering.
On the other hand, the spread of the electron signal (due to diffusion and electronics shaping) makes it difficult to reconstruct the original cluster position.
The large systematic bias in the clustering prevents the use of a Kalman filter~\cite{Fruhwirth:1987fm} to take into account multiple scattering.
Our approach was therefore to focus on the specific geometry that we are looking for: pair vertices.
We developed an original method to efficiently find vertices, and determine the direction of the two associated particles.
Subsequently, vertices of the two projections, $XZ$ and $YZ$, are associated with each other to recover the full 3D description of the vertex.
The method does not yet include a filter to exclude fake vertices.

\subsection{2D Vertex finding}

%We define here a vertex as a region with signal where less than half the field of view contains data. 
We are looking for the point of conversion of a photon into a e$^+$e$^-$ pair.
This means the starting point of two tracks, with an angle strictly inferior to $\pi$.

\begin{figure} [ht]
  \begin{center}
    \includegraphics[width=0.7\textwidth]{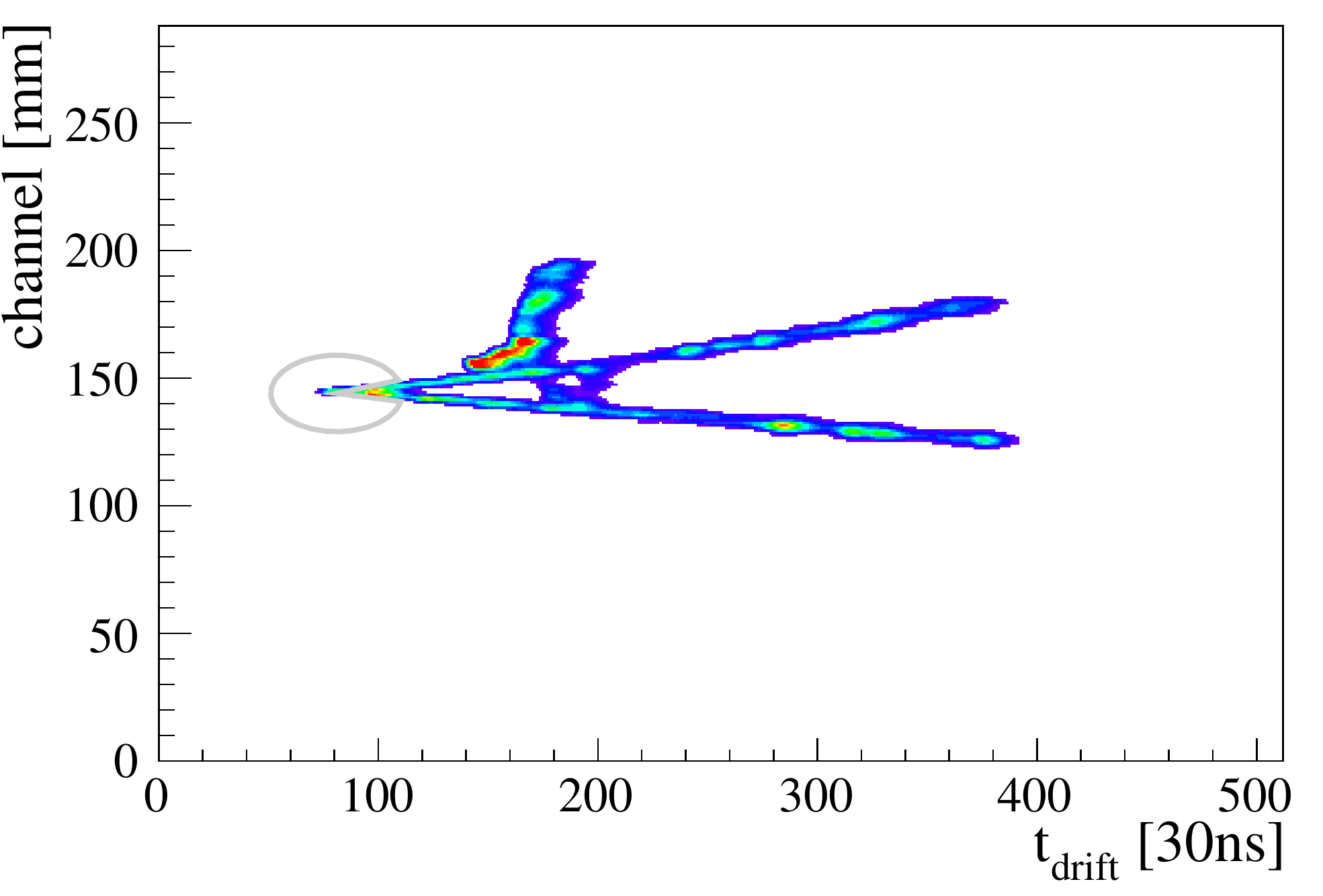}
  \end{center}
  \caption{
    \label{fig:reco:VertexFinding} 
    Vertex finding procedure on a pair conversion event with a $\delta$-electron.
    Circles are drawn around signal regions.
    If the longest arc without signal is larger than half an ellipse, the region is considered as a vertex candidate.
  }
\end{figure}

We scan all the positions containing signal in a grid of $5$\,channels$\times 10$\,time bins (which corresponds to the typical spread of the electron signal). 
We then trace an ellipse around each point and measure the longest arc which does not cross any region containing signal.
If this arc is longer than half the circle, it is identified as a vertex (see Fig.~\ref{fig:reco:VertexFinding}). 
The radius of the ellipse should be larger than the width of the signal (i.e. the electron diffusion and electronics time response), but small enough to avoid being affected by other tracks.
They are taken to be about twice the typical width of the signal.
This procedure relies on the very low noise in the data.

Once a region of interest is found, the position of the vertex is estimated as the centre of gravity of all of the signal within that region.

This method also selects the ends of single tracks and some other regions.
It is necessary, therefore to apply further criteria to select real photon conversion regions.
In the case of the beam data, proximity to the beam position is a sufficient criterion, but for a more generic use, a better description of the event will be needed.

\subsection{2D Vertex fitting}

\begin{figure} [ht]
  \begin{center}
    \includegraphics[width=0.7\textwidth]{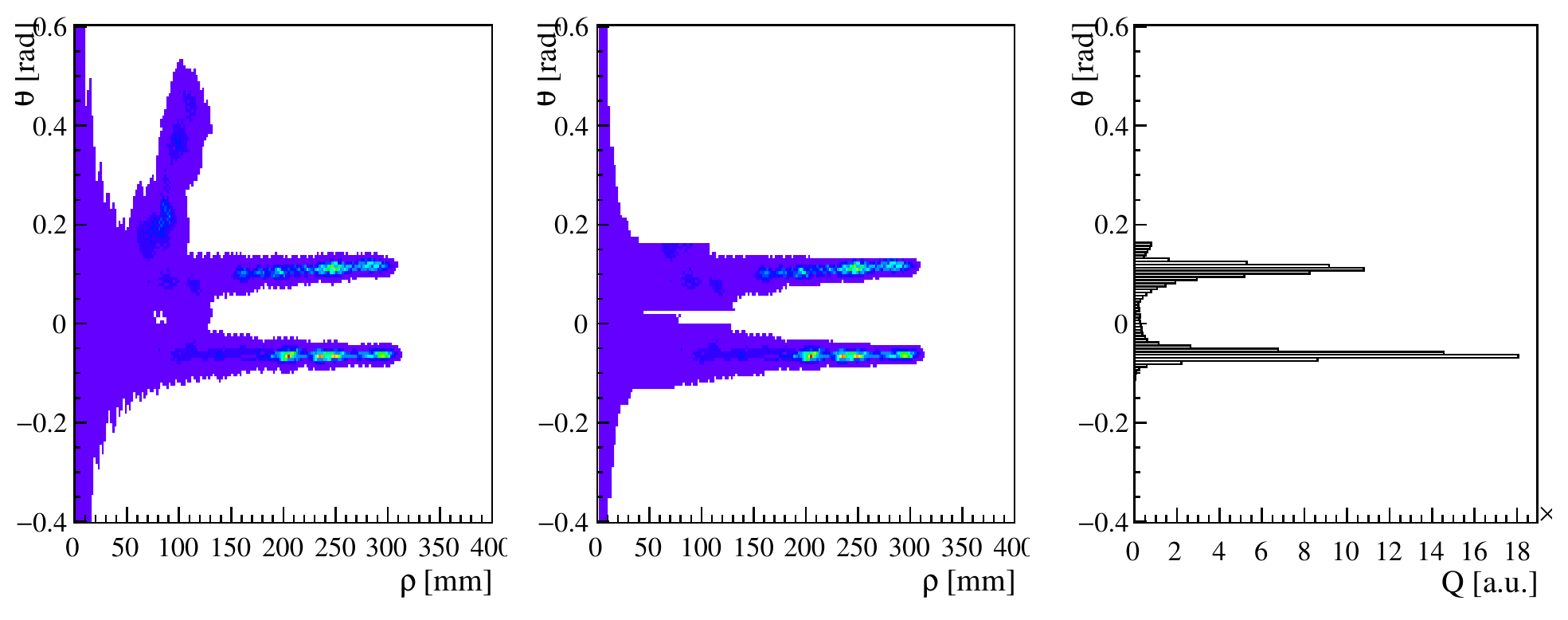}
  \end{center}
  \caption{
    \label{fig:reco:VertexFitting} 
    Vertex fitting procedure.
    Left: polar distribution of charge around the vertex candidate.
%    Middle: only regions connected by a straight line to the vertex are kept (so as to reduce the effect of scattering and background).
    Middle: only regions with nearly continuous signal from the vertex ($\rho=0$) are kept (gaps of one bin are allowed).
    This reduces the effect of scattering and background.
    Most of the signal from the $\delta$-electron is removed.
    Right: angular distribution with two clear peaks corresponding to the two associated tracks. A peak search on the distribution defines the direction of the pair of particles.
  }
\end{figure}

Once a vertex candidate has been found, we can map the signal around it in polar coordinates (Fig.~\ref{fig:reco:VertexFitting}, left).
We only keep the regions with a continuous signal from the vertex (Fig.~\ref{fig:reco:VertexFitting}, middle).

From this, we project the charge distribution as a function of the polar angle around the vertex candidate (Fig.~\ref{fig:reco:VertexFitting}, right).
The peaks of the distribution indicate the direction of the tracks originating from the vertex.

\subsection{Combining 2D vertices for 3D configuration}

\begin{figure} [ht]
  \begin{center} 
    \begin{tabular}{c} 
      \includegraphics[width=0.7\textwidth]{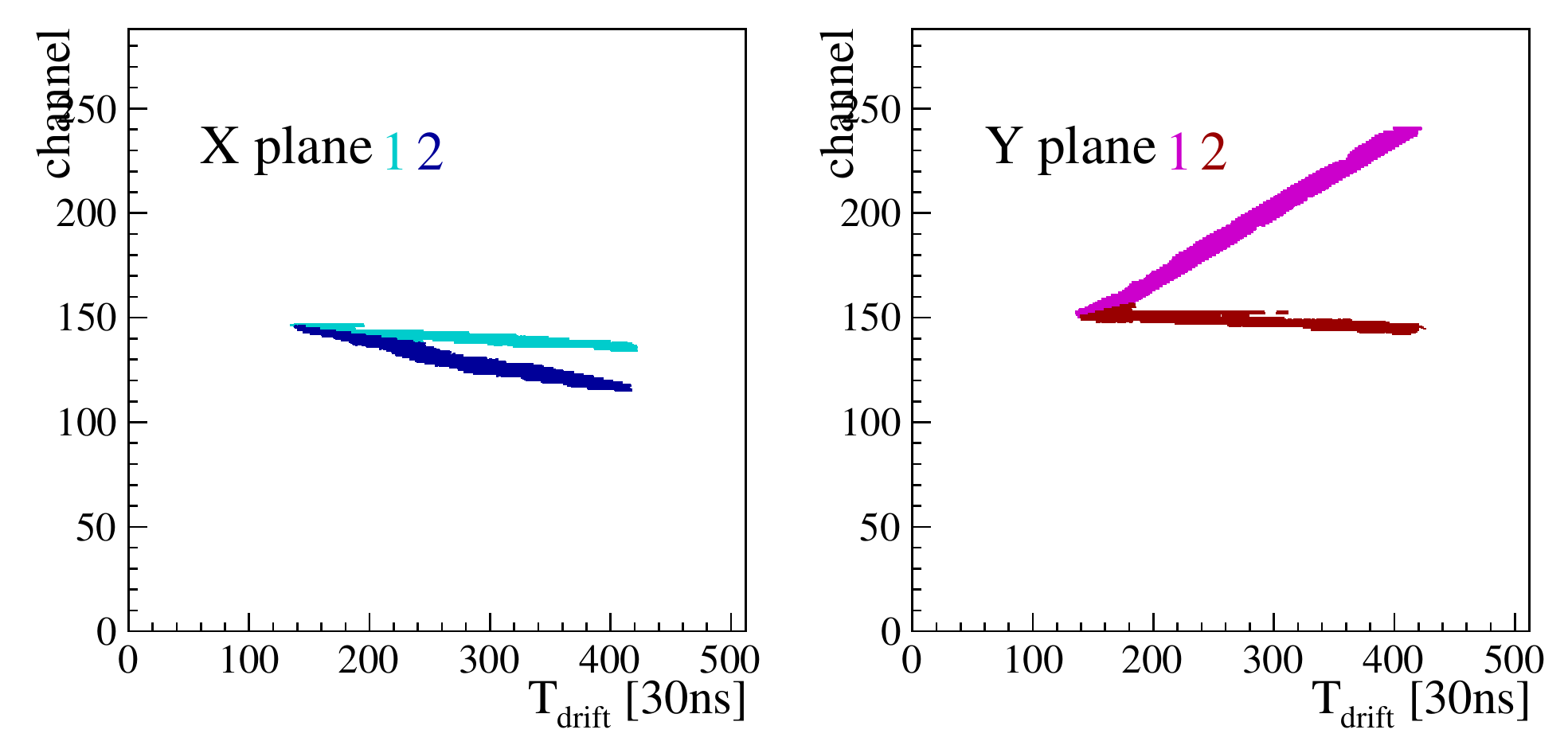} \\
      \includegraphics[width=0.7\textwidth]{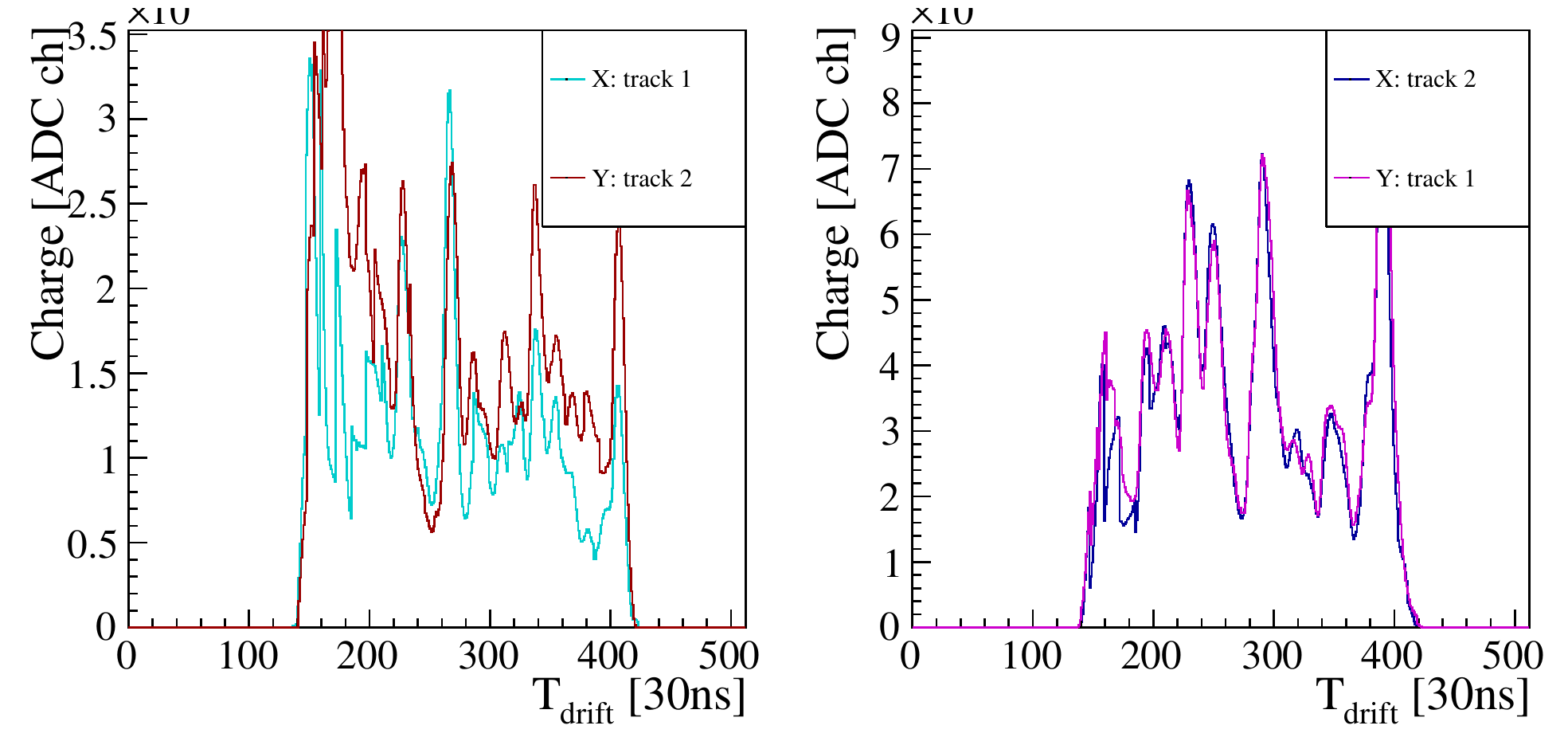}
    \end{tabular}
  \end{center}
  \caption{
    \label{fig:reco:VertexMatching} 
    Procedure for matching the two projections of a vertex.
    Top: definition of the region associated to each of the two particles in the vertex.
    Middle: comparison of the charge profile (in $Z$ direction) of $X$-$Y$ track combinations. 
    The covariance of these distributions is used to select the best match.
  }
\end{figure}

Finally, we associate each $XZ$ projection of a vertex with its $YZ$ projection, and correctly associate the two tracks associated with each vertex, so that we recover a full 3D description of the e$^+$e$^-$ pair.
This is done by comparing the charge profiles of the tracks as a function of the $Z$-coordinate (drift time).
This procedure, called ``matching'' is described in Fig.~\ref{fig:reco:VertexMatching}.

For each 2D vertex, we associate with each of the 2 tracks the signal contained in a region around the track direction.
From this signal, we create the charge profile as a function of $Z$ (or the drift time).
The electron diffusion in the drift volume, and between the GEM plates, create a strong signal correlation at short distances (smaller than $\sim1\milli\metre$).
We know therefore that the charge profile of the same physical object in the $XZ$ and $YZ$ projection should be very similar.

Let's assume we have in the $XZ$ projection a vertex $v_{X}$, with associated tracks $tr^X_1$ and $tr^X_2$, and  in the $YZ$ projection a vertex $v_{Y}$, with associated tracks $tr^Y_1$ and $tr^Y_2$.
Each vertex in the $XZ$ ($YZ$) projection has a position $(X_v,Z^x_v)$ ($(Y_v,Z^Y_v)$).
Each track in the $XZ$ ($YZ$) projection has a direction $(u^X,1)$ ($(u^Y,1)$).
First we require that the $Z$ coordinate of $v_{X}$ and $v_{Y}$ are close ($|Z^X_v-Z^Y_v|<10\milli\metre$.
Then we measure the correlation factor of the charge profiles for each pair of tracks ($tr^X_i$,$tr^Y_j$).
If, for example, ($tr^X_1$,$tr^Y_2$) is the pair with maximum correlation factor, we get a 3D vertex with the following parameters:
\begin{itemize}
\item Position: $\vec{X_{v}} = (X_v,Y_v,(Z^Y_v+Z^Y_v)/2)$)
\item Track A: direction $\vec{u_{A}} = (u^X_1,u^Y_2,1)$ 
\item Track B: direction $\vec{u_{B}} = (u^X_2,u^Y_1,1)$.
\end{itemize}
We have no way of knowing which of the two track ($A$ or $B$) corresponds to the electron or the positron, but this does not affect any of our measurements.

\section{Results on polarimetry} % and angular resolution

We use the vertex information to study the resolution and polarimetry potential of the detector.
A first measurement of this observable was already obtained with a previous version of the event reconstruction (version 2.0).
That version had a much lower efficiency, and the preliminary results were presented and published~\cite{Geerebaert2016}.

\subsection{Reconstruction efficiency}

As explained above, there is, for now, no procedure for selecting vertices based on the event characteristics.
Instead, we use here the fact that we know exactly the source of the photons.
We select therefore the vertices located within the photon beam (which is clearly localised~\cite{Wang:2015thesis}).
The following results are obtained using all of the reconstructed vertices in the data, applying only a space cut around the photon beam.

The work in~\cite{Wang:2015thesis} has shown that the trigger efficiency of the HARPO experiment was above 50\% for all configurations.
For this it identified detection events using the first signal in the TPC.
If this signal is located within the beam region, the event is assumed to be an interaction of the beam with the gas.
Using this definition, we can define the reconstruction efficiency as the fraction of those where a vertex is reconstructed and passes the cut mentioned above.

%We find that the reconstruction efficiency is above 70\%, and reaches up to 90\% depending on the configuration.
We find that the reconstruction efficiency is close to 90\% in the configurations we used.
These results are consistent with the simulation.
%A more detailled study is needed to charaterise the event characteristics which affect this efficiency, although length of the tracks (i.e. the distance to the wall of the TPC) is a major factor.

\subsection{Polarimetry}

%The reconstructed 3D vertices carry the following information:
%\begin{itemize}
%\item the vertex position $\vec{X_{v}}$
%\item the direction of the two tracks $\vec{u_{A}}$ and  $\vec{u_{B}}$
%\end{itemize}
%From this, can compute the azimuthal angle $\omega$. We assume that the incoming photon is in the $Z$~direction so that we have:

We define the azimuthal angle $\omega_{+-}$ of the conversion into an e$^+$e$^-$ pair of a photon propagating along the $Z$ direction (see Fig.~3 in~\cite{Bernard:2013jea}):
\begin{equation}
\omega_{+-} = \arctan\left(u_{e^{+}}^Zu_{e^{-}}^X - u_{e^{+}}^Zu_{e^{-}}^X,u_{e^{+}}^Zu_{e^{-}}^Y - u_{e^{+}}^Zu_{e^{-}}^Y\right)
\end{equation}
The cross-section $\sigma_{conv}$ of conversion for a photon beam with a linear polarisation fraction $P$ is:
\begin{equation}
\sigma_{conv} \propto 1 + A_{QED}P\cos(2\omega_{+-})
\end{equation}
where $A_{QED}$ is the polarisation asymmetry of the conversion process.

\begin{figure}[ht]
  \begin{center}
    \includegraphics[width=0.7\linewidth]{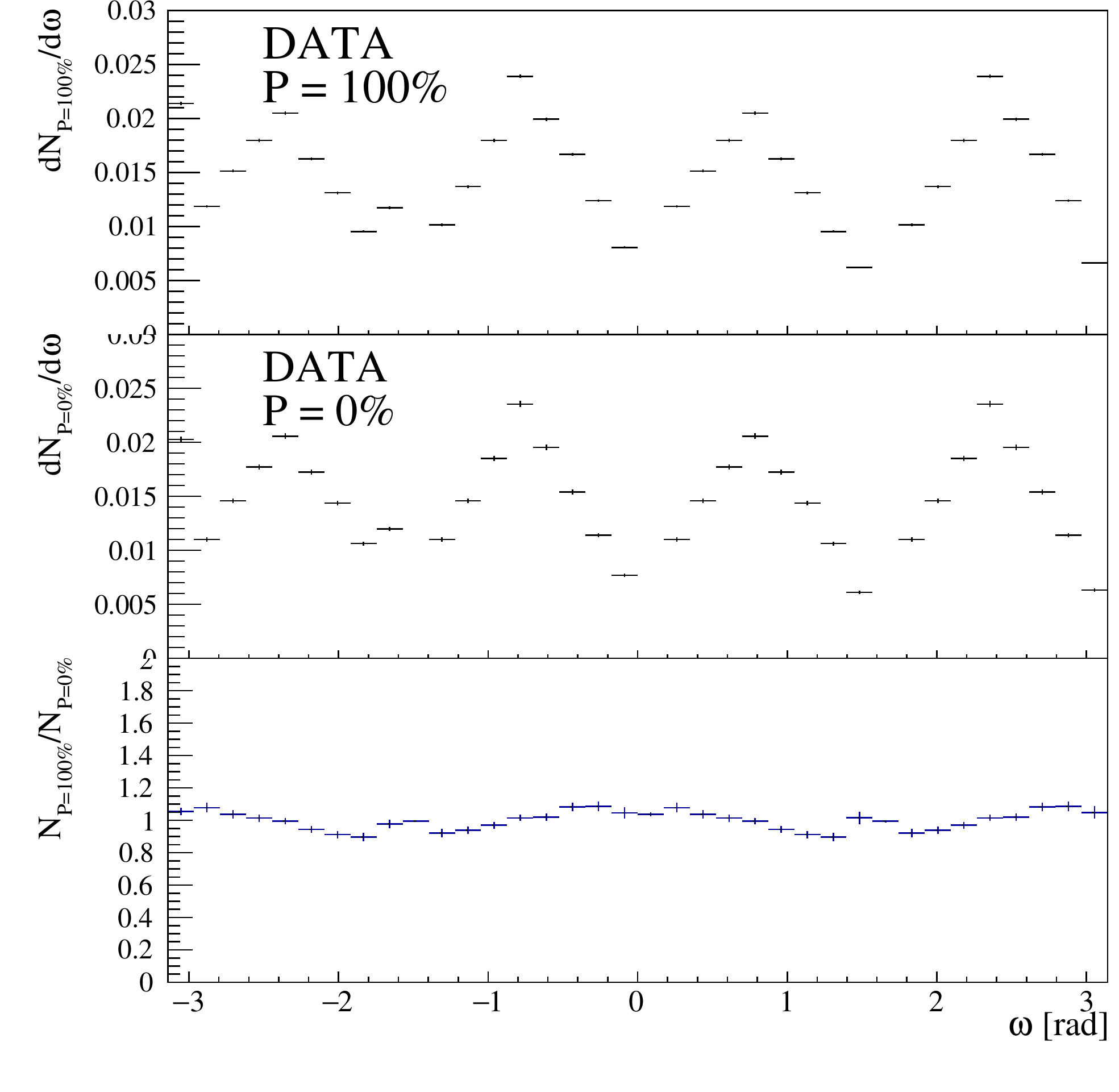}
    \caption{
      Distribution of the azimuthal angle $\omega$ for 11.8\,\mega\electronvolt photons in one angular orientation of the detector.
      Top: fully polarised photon beam. There are large systematic fluctuations.
      Middle: unpolarised photon beam.
      Bottom: ratio of the two above distributions. The systematic effects from the detector geometry cancel out, and the polarisation modulation appears clearly.
      \label{fig:results:omega}
    }
  \end{center}
\end{figure}

Therefore the distribution of the measured azimuthal angle $\omega$, defined as:
\begin{equation}
\omega = \arctan\left(u_{A}^Zu_{B}^X - u_{A}^Zu_{B}^X,u_{A}^Zu_{B}^Y - u_{A}^Zu_{B}^Y\right)
\end{equation}
will follow a distribution:
\begin{equation}
1 + A_{eff}P\cos(2\omega_{+-})
\end{equation}
where $A_{eff}<A_{QED}$ is the effective polarisation asymmetry, taking into account the finite resolution of the detector.
The measured distribution of $\omega$ for 11.8\,\mega\electronvolt photons is shown in Fig.~\ref{fig:results:omega} (top).

We see that there are large systematic fluctuations, mostly related to the cubic geometry of the detector.
%We use data taken in the same detector configuration, but with with an unpolarised beam.
%The resulting $\omega$ distribution in shown in Fig.~\ref{fig:results:omega} (middle).
The resulting $\omega$ distribution with an unpolarised beam is shown in Fig.~\ref{fig:results:omega} (middle).
By taking the ratio of these two distributions, we cancel all of the systematic fluctuations related to the geometry.
The result (Fig.~\ref{fig:results:omega}, bottom) shows clearly the modulation due to the beam polarisation.

\begin{figure} [ht]
  \begin{center}
    \includegraphics[width=0.7\textwidth]{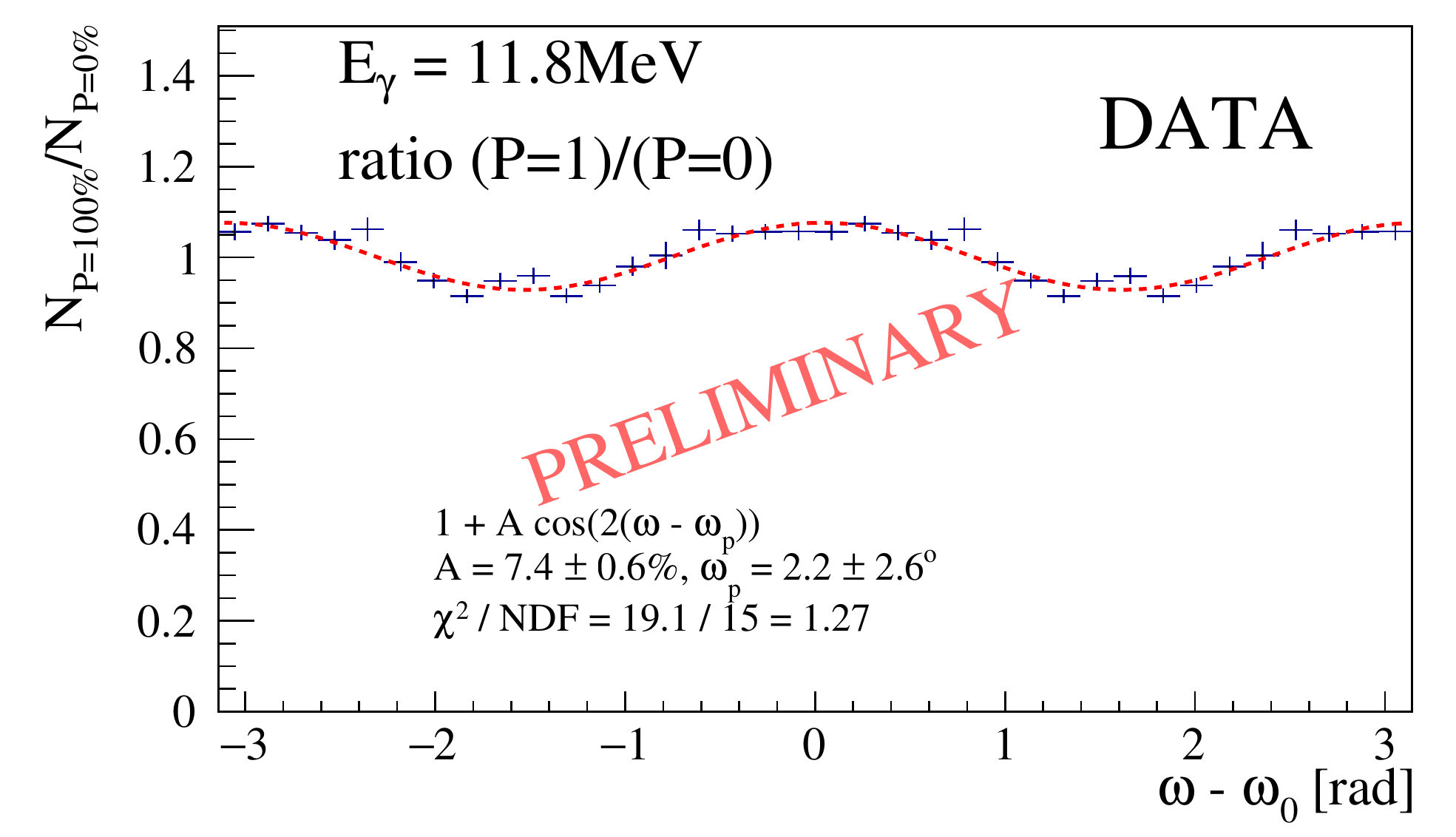}
  \end{center}
  \caption{
    \label{fig:results:pola11MeV} 
    Polarisation modulation for a beam of 11.8\,\mega\electronvolt\ photons.
    The plot is obtained by dividing data taken with a polarised beam and data taken with a non polarised one.
    Most of the systematic errors are cancelled, only statistical errors are shown.
  }
\end{figure}

Finally, we combine the data from the 4 angular configurations of the HARPO detector around the beam axis.
Figure~\ref{fig:results:pola11MeV} shows the resulting distribution.
The cosine modulation is very clear, with an asymmetry parameter $A_{eff}^{beam}=7.4\pm0.6\%$.
%This value of $A$ is slightly smaller than expected in ref~\cite{Bernard:2013jea}.
%A study of the tracking resolution, and of the matching efficiency is needed to explain the value quantitatively.

\begin{figure} [ht]
  \begin{center}
    \includegraphics[width=0.7\textwidth]{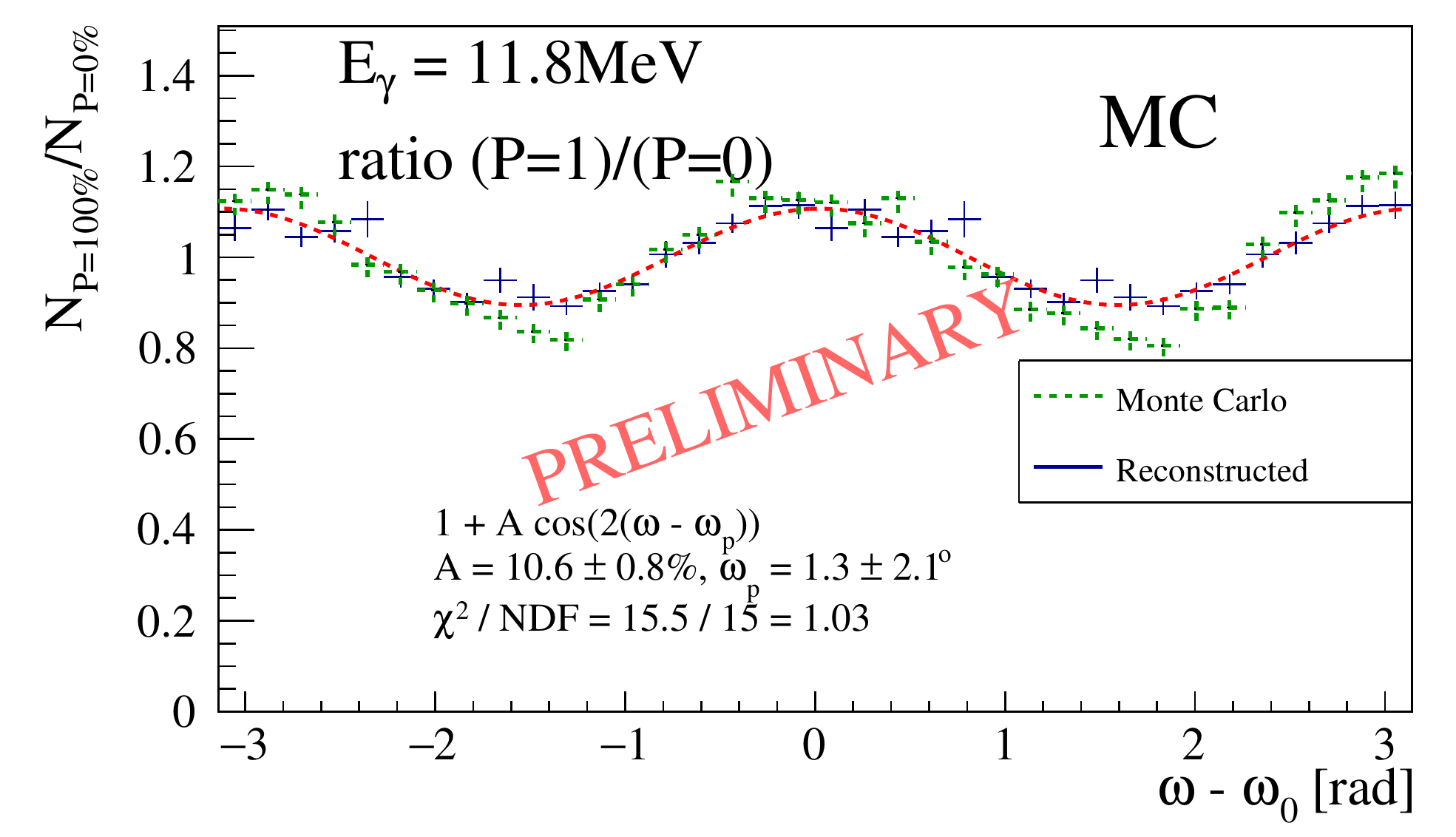}
  \end{center}
  \caption{
    \label{fig:results:pola11MeVsim} 
    Polarisation modulation for a simulated beam of 11.8\,\mega\electronvolt\ photons.
    The plot is obtained by dividing the data taken with a polarised beam by the data taken with a non polarised one.
    Most of the systematic errors are cancelled, only statistical errors are shown.
    In green, the same distribution is made using the exact (i.e. from Monte Carlo) electron and positron direction.
  }
\end{figure}

We apply the same analysis to simulated data.
We simulate pair-conversion events with a conversion point within the beam region of the detector.
No background is included in the simulation.
Figure~\ref{fig:results:pola11MeVsim} shows the polarisation modulation obtained.
We compare this with what would be obtained with ideal tracking (ignoring all detector effects).
We use the exact same procedure, replacing the values of $\vec{u_{A}}$ and  $\vec{u_{B}}$ by the Monte Carlo values.
For 11.8\,\mega\electronvolt\ photons, we find a theoretical asymmetry of $A_{QED}=16.4\pm0.7\%$~\cite{Kotov}, and an effective value of $A_{eff}^{sim}=10.3\pm0.6\%$.
The effective value is significantly higher than the $A_{eff}^{beam}=7.4\pm0.6\%$ from the beam data.
%The detector effects do not influence the asymmetry parameter much.

\subsection{Effects of the tracking resolution on the polarimetry}

\begin{figure} [ht]
  \begin{center}
    \includegraphics[width=0.7\textwidth]{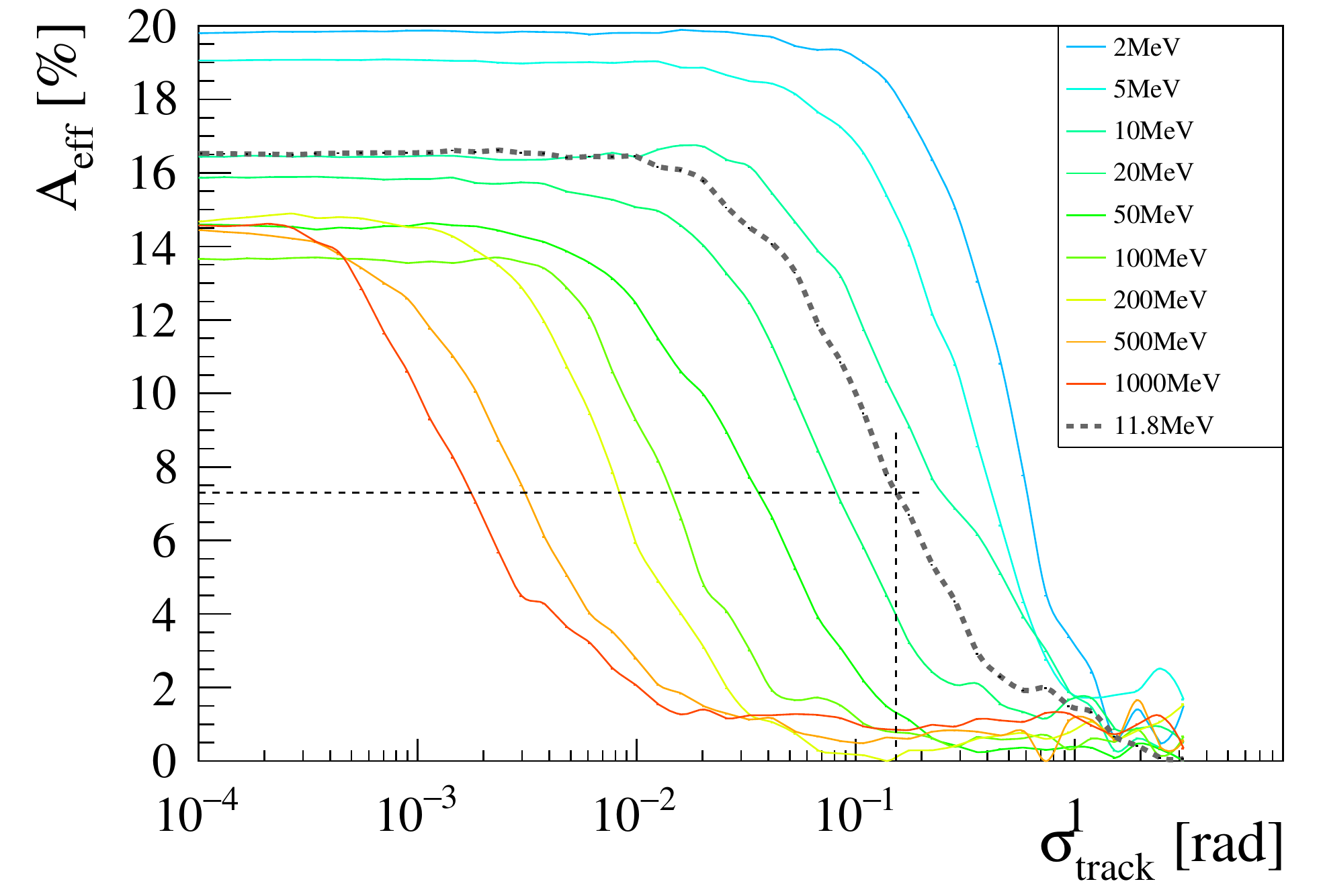}
  \end{center}
  \caption{
    \label{fig:results:AvsResTrack} 
    Effective polarisation asymmetry as a function of tracking resolution for different photon energies.
  }
\end{figure}

%In an ideal case, as shown in Fig.~21 of~\cite{Bernard:2013jea}, the measured polarisation asymmetry $A$ for 11\,\mega\electronvolt photons would be $\sim16\,\%$.
The finite tracking resolution of the detector degrades the effective asymmetry $A_{eff}$.
Figure~\ref{fig:results:AvsResTrack} shows the effective polarisation asymmetry variation with the tracking resolution.
We can see that for each energy, there is a critical resolution where the asymmetry drops.
At higher energies, the opening angle of the e$^+$e$^-$ pair become smaller, so that the resolution of the tracks has more effect on the azimuthal angle, and this critical resolution becomes smaller.

Using the Monte-Carlo information, we can find the track resolution $\sigma_{track}$ (68\,\% containment) in the simulation.
For 11.8\,\mega\electronvolt\ photons, we find $\sigma_{track}=8.79 \pm 0.19\,\degree = 153 \pm 3\,\milli\radian$.
We see in Fig.~\ref{fig:results:AvsResTrack} that the expected corresponding asymmetry is $A_{eff} \approx 7.3\,\%$.
This value is actually smaller than the one obtained after the full simulation and reconstruction.
This shows that the approximation of Gaussian, non-correlated errors on the two tracks is excessive.
In particular, since the two tracks are reconstructed simultaneously in the vertex, their resolution is necessarily correlated.

\subsection{Systematic effects on the polarimetry}

We see a significant difference between the measured asymmetry $A_{beam}=7.4\pm0.6\%$ and the value obtained from simulation $A_{eff}=10.6\pm0.8\%$.
It is clear from Fig.~\ref{fig:results:pola11MeV} that the distribution is not exactly a cosine, with in particular peaks at $\pm\pi/2$.
The method we used cancels all of the systematic effects related to the detector geometry and the reconstruction.
The main remaining source of systematic errors is the beam itself.

There are two background contributions from the beam:
\begin{itemize}
\item 
  Compton scattering of the photons. 
  These can be misidentified as pair conversions due to the saturation effects. 
  The Compton signal is also affected by polarisation.
\item 
  Bremsstrahlung from the electron beam. 
  This gives an unpolarised contribution, with a wide energy distribution. 
  The contribution of bremsstrahlung can vary between runs, due to the variability of the laser intensity, so that the related systematic effects may not completely cancel.
\end{itemize}
These effects have not yet been quantitatively studied.

The results featured in this article show only statistical errors.
A future publication will give final results including systematic uncertainties.

\section{Ongoing work and improvements}

The results shown in these proceedings are still preliminary, using the version 3.0 of the reconstruction and analysis software of the HARPO experiment.
There is ongoing work to improve these results, with in particular:
\begin{itemize}
\item Improvements on the vertex identification and selection.
\item Improvements on the tracking to obtain a better angular resolution and momentum estimation from scattering.
\item Understanding and quantification of the background contamination of the beam data.
\item Calibration of the simulation, and addition of the background, in order to accurately estimate and correct the systematic fluctuations of the azimuthal angle modulation.
\end{itemize}

\section{Conclusions and outlook}

The HARPO detector was completed and used successfully in a photon beam in 2014.
A full simulation of the detector was developed to describe it.
A novel reconstruction algorithm was developed to cope with the unusual event topology.

Even before any optimisation was performed, these tools allowed us to measure the polarisation asymmetry of a 11.8\,\mega\electronvolt\ photon beam with high precision, and with good agreement with simulation.
Some discrepancies remain, and will be addressed by the ongoing analysis and reconstruction developments.
%With further work, the simulation could be used to better understand systematic effects associated with the detector's geometry and correct for them.

The results from the beam data are extremely encouraging and studies are under way to adapt this technology for use as a space telescope.

\section*{ACKNOWLEDGMENTS}       

This work is funded by the French National Research Agency (ANR-13-BS05-0002) and was performed by using NewSUBARU-GACKO (Gamma Collaboration Hutch of Konan University). 

% References
%\bibliography{mybibfile} % bibliography data in report.bib
%\bibliographystyle{ieeetr} % makes bibtex use spiebib.bst

\end{document}